\documentclass[showpacs,preprint,pra]{revtex4}
\usepackage{longtable}

\begin{document}

\title{\bf  Symmetry breaking and structural distortions in 
\\
charged XH$_{4}$( X = C, Si, Ge, Sn and Pb )  molecules} 
\author{ D. Balamurgan, Manoj. K. Harbola  and  R. Prasad }
\affiliation{Department of Physics, Indian Institute of Technology, 
Kanpur 208016, India}
\newpage
\begin{abstract}
We have investigated the ground state structures of neutral and charged 
XH$_{4}$( X=C, Si, Ge, Sn and Pb ) molecules  using the first-principles 
electronic structure methods. The structure of positively 
charged molecules  for X = Si, Ge, Sn and Pb is characterized by a severe 
distortion from tetrahedral structure and an unusual H-H bond while the  
negatively charged molecules get distorted by pushing two  hydrogen atoms 
away from each other. However, CH$_{4}$$^{+}$ and CH$_{4}$$^{-}$ are 
exceptions to this behavior. We  provide an insight into the 
symmetry breaking mechanism  
and unusual H-H bonding using simple electrostatic arguments based on the  
unequal charge distribution on H atoms.  Those charged molecules  having 
unequal charge distribution on H atoms get distorted due to different 
electrostatic forces between the atoms. We show that the 
directionality  and occupation of the highest occupied 
molecular orbital(HOMO) play an important role in creating charge asymmetry 
in these molecules.

\end{abstract}
\maketitle

\section{ Introduction}

Although neutral XH$_{4}$( X= C, Si, Ge, Sn and Pb ) molecules have tetrahedral 
symmetry, charging these molecules breaks this  symmetry.  The symmetry 
breaking in general is associated  with a large structural distortion and in 
many cases  unusual H-H bonding arrangement.  Some of these clusters have been 
investigated earlier and the  symmetry breaking has been attributed to the 
Jahn-Teller effect.  In the Jahn-Teller 
effect\cite{BersukerCM01v101,solidv20,Englman},  electrons 
occupying a degenerate level favor a 
low symmetric structure by  lifting the  degeneracy.  For example, the  
highest occupied molecular orbital(HOMO) of tetrahedral 
SiH$_{4}$ is triply degenerate. When an electron 
is removed from SiH$_{4}$, its structure gets distorted drastically due to the 
Jahn-Teller effect\cite{FreyJCP88v89,ProftCPL96v262,KudoCP88v122,
CaballolCPL86v130,PorterJCP01v114}.  

The structure of SiH$_{4}$$^{+}$ has been investigated by a large number of 
workers\cite{FreyJCP88v89,ProftCPL96v262,KudoCP88v122,
CaballolCPL86v130,PorterJCP01v114} and  
it took almost a decade to  settle its ground state 
structure. Now it is generally agreed that the ground state structure of 
SiH$_{4}$$^{+}$ has C$_{s}$ symmetry.  Similarly, the   
structure of CH$_{4}$$^{+}$ has been studied  
extensively\cite{PorterJCP01v114,WetmoreJCP99v110,VagerPRL86v57,
SignorellJCP99v110,VegaJMOL85v123,BoydJCP91v94,SignorellJES00v108,
TakeshitaJCP87v86,FreyJCP88v88}, though there is 
still ambiguity regarding its ground state structure. The calculations 
done by the density functional theory( DFT )  methods show that the ground 
state structure of CH$_{4}$$^{+}$ 
has D$_{2d}$ 
symmetry\cite{PorterJCP01v114,WetmoreJCP99v110}, while  the 
Hartree-Fock( HF ) 
and post HF calculations show   
the ground state structure as 
C$_{2v}$\cite{WetmoreJCP99v110,FreyJCP88v88}. The lowering of 
tetrahedral symmetry 
of CH$_{4}$$^{+}$ and SiH$_{4}$$^{+}$ has been  attributed to  
the Jahn-Teller effect.  However, 
the structure of 
CH$_{4}$$^{+}$  is quite different from that of SiH$_{4}$$^{+}$. It is 
difficult to understand the difference between the CH$_{4}$$^{+}$ 
and SiH$_{4}$$^{+}$ distortions only from the Jahn-Teller effect. This is 
because the Jahn-Teller theorem only 
suggests the existence of  distortion and  says nothing regarding the 
nature and magnitude of the distortion. 

In this paper we focus on the ground state structures of charged XH$_{4}$
molecules using the first-principles electronic structure methods
and provide an  insight  
into the symmetry breaking mechanism. We find that
upon charging most of the molecules undergo large structural distortion 
which can be
attributed to the Jahn-Teller effect.  Further, the 
nature of distortions of positively charged
molecules is different from that of negatively charged 
molecules.  Interestingly, most of the positively charged molecules have 
an unusual H-H bond and appear like transition states of unstable
XH$_{2}$$^{+}$ and H$_{2}$ complex.  We try to understand the distortions,
unusual H-H bonding and stability of these molecules.  Vibrational analysis 
shows that these
molecules are not transition states. The calculated fragmentation energy
indicates
that the stability of  XH$_{4}$$^{+}$ molecules decreases from
CH$_{4}$$^{+}$ to PbH$_{4}$$^{+}$.

Using various 
exchange-correlation functionals such as local density approximation(LDA),   
local spin density approximation(LSDA), gradient corrected 
approximation(GGA) and spin-polarized GGA(GGA-SP) we have investigated  
the effect of different occupations in  charged 
tetrahedral XH$_{4}$ molecules. Our calculations show that the configuration
corresponding to integer occupancy and asymmetrical charge density leads
to the lowest energy.  We provide simple arguments in terms
of charge asymmetry and electrostatic interaction between atoms
which  explain  the nature of distortions in charged XH$_{4}$
molecules. For example, we are able to explain why the nature of distortion
of SiH$_{4}$$^{+}$ is different from CH$_{4}$$^{+}$ and why the positively
charged molecules distort differently from the negatively charged
molecules. We study the structural distortions of SiH$_{4}$$^{+}$ in
detail using conjugate gradient method and show that the
evolution of the structure is governed by the electrostatic interaction
between the atoms. Our investigations reveal that 
the distribution of
charges on H atoms plays an important role in distorting the charged
molecules. If the charges on H atoms are not equal, the structure
gets distorted due to  different electrostatic forces on the
atoms.  We also discuss how the charge asymmetry is created
in these molecules by the directionality  and occupation of
the HOMO.

The plan of the paper is as follows. In section \textbf{II}
we give computational details of the present work.  Section \textbf{III}  
contains  results and discussion and finally in section \textbf{IV} 
we give our conclusions.

\section{ Computational details} 

We have investigated the structure of neutral and charged XH$_{4}$  
using Vienna \emph{ab-initio} simulation 
package( VASP )\cite{KressePRB96v54} and Gaussian98\cite{Gaussian98} 
package.  We have cross checked our calculations using both packages and  
find that the results obtained by both methods are in good agreement. 
A number of calculations were done using the unrestricted 
Hartree-Fock( UHF ) and  DFT   methods using the Gaussian98 package. In the  
DFT calculations  we have used several 
 exchange-correlation functionals such
as   Slater's exchange with the local correlation 
functional of Perdew( SPL ),   Becke's exchange which includes gradient 
density with gradient corrected correlation functional( BPW91) and 
Becke's 3 parameter hybrid 
functional( B3LYP )\cite{Gaussian98,Foresman}.  In 
these calculations we have  employed    
6-311g** basis set for molecules having 
C and Si atoms and LanL2DZ  basis set with effective core potential 
for other molecules.  The 
plane wave ultrasoft pseudopotential 
calculations\cite{KressePRB96v54} were done with  the 
local density 
approximation( LDA )\cite{CeperleyPRL80v45,PerdewPRB81v23} 
employing a 
simple cubic supercell of 20$\AA$.  The number of plane 
waves used in the calculation is set by energy cut-off( Ecut). We 
have used  287 eV Ecut for neutral and charged CH$_{4}$ molecules 
and 150 eV Ecut for other molecules.  The use of supercell technique for
charged systems causes electrostatic interaction within the images of
supercells\cite{NeugebauerPRB92v46,MakovPRB95v51}. Calculation
of charged systems are efficiently handled in VASP by applying a background 
charge to maintain the charge neutrality and by adding dipole  and
quadrupole moment corrections\cite{KressePRB96v54}.  The structure 
and ionization potential of SiH$_{4}$$^{+}$  predicted by using this procedure
are in close agreement with those of earlier calculations
based on  the HF, MP2 and LSDA 
methods\cite{KudoCP88v122,PorterJCP01v114}.

\section{ Results and Discussion} 
\subsection{  Ground state structures} 

Fig. 1 shows the ground state structures of XH$_{4}$ molecules under 
neutral and charged states using the Gaussian98 package with BPW91 
functional.  We first discuss the ground state structures of the positively 
charged molecules. We see from the figure that the tetrahedral symmetry is 
broken when  XH$_{4}$ molecules are ionized.  The 
structure of CH$_{4}$$^{+}$  has D$_{2d}$ symmetry, while other 
positively charged molecules have C$_{s}$ symmetry. We have investigated 
various possible symmetries of XH$_{4}$$^{+}$ molecules  
and our results are summarized in Table. I. We see from the 
table that C$_{2v}$ and D$_{2d}$ structures of  CH$_{4}$$^{+}$ 
are very close in energy and differ only by 0.11 eV. The other positively 
charged 
molecules also have C$_{2v}$ as the next higher energy structure.  We 
could not find  local minima of C$_{3v}$ and C$_{s}$ symmetries for 
CH$_{4}$$^{+}$ and C$_{2v}$ symmetry for SiH$_{4}$$^{+}$ at the BPW91 level. 
These missing structures of CH$_{4}$$^{+}$ and 
SiH$_{4}$$^{+}$  molecules in our calculations have been 
studied earlier using different exchange-correlation potentials and 
different methods\cite{FreyJCP88v89,ProftCPL96v262,KudoCP88v122,
WetmoreJCP99v110,TakeshitaJCP87v86,FreyJCP88v88}  which show 
that these are not the 
ground state structures.

It is interesting to 
note that in the C$_{s}$ symmetry, the positively charged molecules have a 
structure which is  severely distorted from the tetrahedral 
structure. Furthermore, there is an unusual bonding between  two hydrogen 
atoms. This bonding is confirmed by the charge density calculation as shown 
in Fig. 2 which shows large electron density between the two hydrogen 
atoms of SiH$_{4}$$^{+}$. To ascertain that this is not an artifact of the 
BPW91 calculation, we have re-optimized the ground state structures using 
the HF and DFT methods  
with different exchange correlation functionals such as 
SPL  and B3LYP. The 
H-H bond lengths obtained  from various methods are essentially the 
same.  This kind of 
over-coordination of hydrogen has also been observed in 
CH$_{5}$$^{+}$\cite{MarxSci99v284,MarxAng97v36}, 
SiH$_{5}$$^{+}$\cite{BooJCP95v103} and 
small hydrogenated silicon clusters\cite{BalaPRB01v64}.  In Fig. 3, we 
have shown the total energy curve of SiH$_{4}$$^{+}$, 
H$_{2}$ and H$_{2}$$^{+}$ as a function of H-H distance. The total 
energy of SiH$_{4}$$^{+}$ is calculated by keeping Si-H bond length 
fixed but varying H-H bond distance.  From the figure we see that the 
curvature of SiH$_{4}$$^{+}$ curve is smaller than H$_{2}$ and larger than 
H$_{2}$$^{+}$ curves. This shows H-H bond in SiH$_{4}$$^{+}$ is 
stronger than H$_{2}$$^{+}$ but weaker than H$_{2}$.  Vibrational 
analysis also indicates  
that the H-H bond is sufficiently strong in these molecules.  We see 
from  Table. II that the force constant for   
H-H stretch mode of H-H pair is smaller than 
H$_{2}$ and larger than H$_{2}$$^{+}$ which  
implies that the strength of H-H bond 
is weaker than H$_{2}$ and stronger than H$_{2}$$^{+}$. 

For the ground state structures of the negatively charged 
molecules,  we note from Fig. 1 that the  distortions of negatively 
charged molecules 
are different from those of positively charged molecules.  XH$_{4}$$^{-}$ 
molecules have C$_{2v}$ symmetry except  
CH$_{4}$$^{-}$ which has tetrahedral symmetry. The two hydrogen atoms of 
XH$_{4}$$^{-}$ molecules 
are pushed away from each other which is in contrast to the H-H bond 
of XH$_{4}$$^{+}$ molecules with C$_{s}$ symmetry.

\subsection{ Stability}

One may argue that the structure of  XH$_{4}$$^{+}$ molecules
with C$_{s}$ symmetry appears more like XH$_{2}$$^{+}$ and H$_{2}$
complex and hence unstable.  Frey and
Davidson\cite{FreyJCP88v89} studied  SiH$_{4}$$^{+}$ using
CI singles root calculation and showed that C$_{s}$ structure is more 
stable than C$_{2v}$  against fragmentation  into  SiH$_{2}$$^{+}$ 
and H$_{2}$.   We have calculated 
the fragmentation energy which is the energy  required to 
fragment a XH$_{4}$$^{+}$ molecule into its binary 
products and has been calculated by taking the total energy difference 
between the XH$_{4}$$^{+}$ and the possible binary products.  Of all the 
possible 
channels, the fragmentation   
of XH$_{4}$$^{+}$ into XH$_{2}$$^{+}$ and H$_{2}$ requires least energy.  We 
give  the  fragmentation energies of XH$_{4}$$^{+}$ molecules 
for this process in Table. III. We see from the table 
that the energy required for the fragmentation of XH$_{4}$$^{+}$ 
into XH$_{2}$$^{+}$ and H$_{2}$ 
decreases for X=C to Pb.  This means that the  stability of XH$_{4}$$^{+}$ 
molecules decreases from C to Pb molecules.

To examine the stability further we have done vibrational 
analysis of these molecules.  In Table. IV  we present 
the vibrational frequencies of charged and neutral XH$_{4}$ molecules.  Our 
calculated vibrational frequencies 880, 950, 2175 and 2192cm$^{-1}$ of 
SiH$_{4}$ are in good agreement with those of  experimentally obtained
values\cite{CoatsJMol94v320} of 914, 953, 2189 and 
2267cm$^{-1}$.  We  see from 
the table  that PbH$_{4}$$^{+}$  has a negative 
frequency and other molecules  have  only positive frequencies. Further, 
we see that one frequency is very
low for  C$_{s}$ symmetric XH$_{4}$$^{+}$ molecules and 
decreases from SiH$_{4}$$^{+}$ to GeH$_{4}$$^{+}$ and become
negative for PbH$_{4}$$^{+}$.  

The positive frequencies of a molecule 
indicate that the structure 
corresponds to an energy minimum. If one  frequency is negative, the structure 
may be a transition state of the molecule\cite{Foresman}.  Since the 
frequency 
calculation is highly dependent on the method, the small negative frequency of 
PbH$_{4}$$^{+}$ may be an artifact of the method. To clarify this 
we have done vibrational analysis on  re-optimized structures of 
C$_{s}$ symmetric XH$_{4}$$^{+}$ molecules using various methods. Calculation 
with the 
SPL exchange-correlation functional shows that one frequency of 
PbH$_{4}$$^{+}$  which was negative in the calculation 
with BPW91 functional is positive but near zero. 
Similarly, calculation with the UHF 
method also shows 
that all the frequencies of PbH$_{4}$$^{+}$ are positive. Irrespective of the 
sign of calculated  frequencies,  all these methods show that one 
vibrational mode of C$_{s}$ symmetric XH$_{4}$$^{+}$ molecules has a low 
frequency.  Also we find that the force constant corresponding 
 to this mode is nearly zero, which implies that the H atoms in these mode  
are moving on  nearly flat potential energy surfaces. 

We  see from Fig. 1 that the  mirror symmetry plane
of C$_{s}$ structure is formed by the H-H pair and X atom. 
The vibrational mode with low frequency corresponds
to  the vibration of
H-H pair about the mirror symmetry plane of  C$_{s}$ structure.  When the H-H 
pair vibrates about the plane, one 
atom of H-H pair  moves in and the other moves out of the mirror 
plane without change in the H-H distance.  Since the 
force constant 
is nearly zero for this vibration,  the motion of H-H pair is 
more like an internal rotation about an axis which is 
perpendicular to the H-H bond and passing through the 
X atom. This internal rotation of H-H pair 
in XH$_{4}$$^{+}$ is similar to the internal rotation of O-H in 
CH$_{3}$OH molecule\cite{AyalaJCP98v108,QuadeJMS98v188,DuanCP02v280}.  Thus the 
low frequency mode in the present case does not indicate a transition state. 

\subsection{ Effect of different occupations} 

Before proceeding further, we would like to resolve an interesting 
question regarding the occupation of degenerate states in charged 
tetrahedral XH$_{4}$ clusters.  Let us ionize a charged XH$_{4}$ 
cluster keeping its tetrahedral 
structure intact.  In this structure the HOMO is  
triply degenerate and is occupied by only 5 electrons.  Hence, the HOMO of 
XH$_{4}$$^{+}$ is not completely filled.  In such case of   
incompletely filled degenerate systems, 
there are many possible ways to fill the levels. The question is 
what is the right way of filling the levels and which distribution  
leads to the lowest energy. The usual practice in such situation is 
to fill all levels with fractional number of electrons, thereby keeping 
the symmetry intact.  

 To  resolve  the issue of occupation  
we have performed paramagnetic and spin-polarized calculations on 
tetrahedral SiH$_{4}$$^{+}$ and CH$_{4}$$^{+}$ using VASP package with 
local and non-local functionals for various possible electronic 
distributions.  In paramagnetic calculations 
we have chosen two possible electronic distributions.  The 
configuration with integer( 2,2,1 ) occupation
corresponds to asymmetric charge density and the other with
fractional( $\frac{5}{3}$, $\frac{5}{3}$, $\frac{5}{3}$ ) occupation
corresponds to symmetric charge density.  Note that the spin multiplicity of 
SiH$_{4}$$^{+}$ and CH$_{4}$$^{+}$ is unity. The multiplicity is
preserved in charge asymmetric configuration whereas it is
zero for charge symmetric configuration. In spin-polarized calculations
we have chosen three possible electronic configurations.  The
configuration  with $\frac{5}{3}$$\uparrow$,
$\frac{5}{3}$$\uparrow$, $\frac{5}{3}$$\uparrow$
( $\frac{5}{3}$$\downarrow$,
$\frac{5}{3}$$\downarrow$, $\frac{5}{3}$$\downarrow$ ) occupation 
corresponds to charge symmetric density and multiplicity zero.  Another 
configuration with 
1$\uparrow$, 1$\uparrow$, 1$\uparrow$( $\frac{2}{3}$$\downarrow$,
$\frac{2}{3}$$\downarrow$, $\frac{2}{3}$$\downarrow$) occupation
also corresponds to
symmetric charge density  but multiplicity one.  The other
configuration with  1$\uparrow$, 1$\uparrow$, 1$\uparrow$( 1$\downarrow$,
1$\downarrow$, 0$\downarrow$) occupation corresponds to
asymmetric charge density and  multiplicity one.  We note that in 
all these configurations, the Kohn-Sham wave function is a single 
slater determinant. Hence the energy of the multiplet can be 
calculated directly using von Barth scheme\cite{BarthPRA79v20}.  The 
calculated 
total energy of tetrahedral SiH$_{4}$$^{+}$ and CH$_{4}$$^{+}$
molecules are summarized in Table. V. 

From the table we see that
paramagnetic calculations with local( LDA )  and non-local( GGA )
functionals  do not give much difference between total energies of the 
molecules for  symmetric and asymmetric charge density.  On the other hand, 
the spin polarized calculations with 
local functional( LSD )
favors charge symmetry for SiH$_{4}$$^{+}$ and charge asymmetry for
CH$_{4}$$^{+}$. However, for both the molecules LSD calculations show that 
the symmetric charge density which has multiplicity of zero is 
highly unfavored. We note that in spin
polarized calculations the exchange interaction is taken care of 
more accurately than in paramagnetic calculations. Hence the unphysical
zero multiplicity is strictly not favored
by spin polarized calculations. But LSD does not give consistent results
for multiplicity preserved symmetric and
asymmetric charge density of SiH$_{4}$$^{+}$ and CH$_{4}$$^{+}$ 
molecules.  This is a well known shortcoming of the 
LSD in the context of atomic calculations\cite{JanakPRB81v23}. This  
deficiency of LSD  is removed by the gradient corrected 
functionals\cite{KutzlerPRL87v59}.  We have therefore done spin 
polarized GGA( GGA-SP) calculations on SiH$_{4}$$^{+}$ and CH$_{4}$$^{+}$ 
molecules.  Our calculations with  
GGA-SP functional give the consistent result that the
asymmetric charge density is
the ground state configuration for both SiH$_{4}$$^{+}$ and CH$_{4}$$^{+}$
molecules in tetrahedral structure.

Apart from the
total energy calculation, one more important issue is the calculation
of forces in such systems.  We find that
enforcing charge symmetry with fractional  
$\frac{5}{3}$, $\frac{5}{3}$, $\frac{5}{3}$ occupation 
in the degenerate level of
XH$_{4}$$^{+}$ in LDA calculations does not distort the
tetrahedral structure.  This is in contrast with the 
structural relaxation of tetrahedral XH$_{4}$$^{+}$ 
with 2,2,1 occupation of degenerate level where the molecules 
results in distorted structure.  Although LDA does not distinguish clearly
the charge symmetric and asymmetric configurations, the calculated
forces are different for the two configurations. This  indicates that by 
enforcing symmetric charge
distribution, the forces on atoms of XH$_{4}$$^{+}$ molecules are
incorrectly estimated.  It was also noted 
by Janak and Williams\cite{JanakPRB81v23} that enforcing
spherical charge distribution on atoms can give rise to errors in
Hellman-Feynman forces.  Therefore, our results clearly indicate that the 
electron charge density of  tetrahedral XH$_{4}$$^{+}$  is asymmetric.

\subsection{ Structural evolution} 

To gain insight into the mechanism of structural distortion in SiH$_{4}$$^{+}$ 
we look at the evolution of SiH$_{4}$$^{+}$ structure by using 
the VASP package with LDA functional 
and  the conjugate gradient(CG) method\cite{Steep}.  We first 
ionize SiH$_{4}$ by holding  atoms to obtain 
tetrahedral SiH$_{4}$$^{+}$ and then relax the structure.  We find that 
with 64 steps the system  rolls down from the initial tetrahedral to a well 
converged ground state structure. Each CG step corresponds to a particular 
geometrical configuration. The bond length between silicon atom and the 
distance between the hydrogen atoms as a function of CG steps are shown 
in Fig. 4(a). Initially one H atom moves away from the silicon atom and attains
maximum Si-H bond length of 1.92$\AA$ at 8th step. Then this repelled 
H atom moves closer to the nearest hydrogen from 13th step 
which can be seen from the H-H distance shown in Fig. 4(b). The two 
hydrogen atoms come very close to each other to a distance of 
0.84$\AA$ and form an H-H pair.  At this stage the structure of SiH$_{4}$$^{+}$ 
is C$_{2v}$. After  23rd step, this H-H pair gets rotated towards 
the other two hydrogen atoms without changing the H-H pair distance 
which finally results in C$_{s}$ symmetric  SiH$_{4}$$^{+}$.   

The summation of  occupied Kohn-Sham orbital 
energies  as a function of CG step is shown in the Fig. 4(c). From the 
figure we can see that the sum  
increases initially up to the 8th step  and then decreases  
enormously  between 13th and 23rd step. Then it  fluctuates between 24 and 
40th step  and eventually  converges to a constant value. The 
figure shows that the structural changes occurring from 13 
and 23 are due to lowering in the orbital energies. Note that there is 
a minimum at 23rd step. As mentioned earlier at this CG step  the structure 
of SiH$_{4}$$^{+}$ is C$_{2v}$ with  the H-H  bond.  In a sense, the sum of 
orbital energies shown in the figure is similar to the Walsh 
diagram\cite{Walsh} where 
one takes into account only the sum over orbital energies in predicting the 
structure. In Walsh diagram, each orbital energy is plotted as function of 
bond angle of a molecule. Then the minimum of the sum  helps in predicting 
the structure. Thus in the present case  the Walsh diagram would predict 
the symmetry of  SiH$_{4}$$^{+}$ as C$_{2v}$ symmetry  and not  C$_{s}$,  
which is the correct symmetry.

\subsection{ Role of electrostatic interaction}

We now show that the distortion and formation of H-H bond can be easily 
understood from  arguments based on electrostatic 
interaction between atoms of the  molecule. We first try to understand the 
structural evolution of SiH$_{4}$$^{+}$  as discussed in the previous 
subsection. Let us first consider neutral SiH$_{4}$ molecule. Since H is 
more electronegative\cite{RoblesJACS84v106}than Si, the H atoms in 
SiH$_{4}$ has a small negative charge while Si will have a small positive 
charge. This is also shown by  population 
analysis\cite{Gaussian98,Foresman,Pop} 
according to which  each hydrogen atom holds a charge of -0.12e and silicon 
+0.52e. When the molecule is ionized to SiH$_{4}$$^{+}$ holding its 
atoms fixed,  population analysis shows that  the hydrogen atoms have  
-0.12e, -0.02e, +0.06e and +0.30e  charge 
and silicon  +0.77e charge. Interestingly, hydrogen 
atoms of  tetrahedral SiH$_{4}$$^{+}$ do not have equal charge although the 
structure is symmetric.  One hydrogen atom has  more positive charge than 
the others and gets repelled by silicon atom which is also positively 
charged. This hydrogen moves away from the silicon to reduce the 
electrostatic repulsion and is attracted by the nearby hydrogen atom which 
has more electrons and forms an H-H bond. This H-H pair is positively 
charged and hence attracted  by the nearby two hydrogen atoms  which results in 
rotation of the pair towards the other two hydrogen atoms.

Since the electronegativities of Ge, Sn and Pb atoms are lower than 
that of hydrogen\cite{RoblesJACS84v106}, GeH$_{4}$$^{+}$, SnH$_{4}$$^{+}$ 
and PbH$_{4}$$^{+}$ 
have  distortions and bonds similar to  that of SiH$_{4}$$^{+}$.  However, as 
noted earlier,  the structural distortion of CH$_{4}$$^{+}$ is different from 
that of SiH$_{4}$$^{+}$; two hydrogen atoms of CH$_{4}$$^{+}$ move away 
from each other which is in contrast with the formation of H-H bond 
in SiH$_{4}$$^{+}$. We now show that the same arguments based on 
electronegativity  and electrostatic interaction also explain the 
structural distortion of CH$_{4}$$^{+}$. Since carbon has larger 
electronegativity than H atom, carbon in CH$_{4}$ will have negative charge 
while hydrogen will have positive charge. This is confirmed by  population 
analysis which shows that a charge of -0.82e on carbon and +0.20e on each 
hydrogen atom. Ionizing  CH$_{4}$ in the tetrahedral symmetric structure 
results in  +0.20e, +0.23, +0.42e and +0.55e charge on  H atoms  and 
-0.40e charge on carbon. We  see that  the charge distribution is 
asymmetric and two hydrogen atoms are more positively charged than 
others. The two  H atoms  repel each other  and open  H-C-H bond angle as 
seen in Fig. 1.

Now we discuss the distortion of negatively charged molecules and explain it 
on the basis of the above arguments. We see from Fig. 1 that 
CH$_{4}$$^{-}$  has the tetrahedral symmetry while other negatively charged 
molecules have  C$_{2v}$ symmetry.  Putting an electron on CH$_{4}$ results 
in -1.00e on carbon  and  nearly zero electronic charge on H atoms. Because 
of this symmetrical charge 
distribution on hydrogen atoms, the forces on H atoms are radial and the 
structure of CH$_{4}$$^{-}$ remain tetrahedral. In contrast, in other 
molecules the charge on H atoms are not equal. For example, in 
SiH$_{4}$$^{-}$ population analysis shows a charge distribution of -0.14e, 
-0.15e, -0.16e and -0.37e on H atoms   while silicon holds 
a charge of -0.18e. We can see that the H atom which is more negative gets 
repelled by other atoms. As a result, one of the  H-Si-H bond angle 
opens up with the corresponding expansions in Si-H bond lengths. The trend 
in the electron distribution of GeH$_{4}$$^{-}$, SnH$_{4}$$^{-}$ and 
PbH$_{4}$$^{-}$ is similar to that of SiH$_{4}$$^{-}$. Therefore, the 
distortion of these molecules are similar to that of SiH$_{4}$$^{-}$ and can 
be explained using the same reasoning.

\subsection{ Charge asymmetry}

It is interesting to note that in most of the charged molecules the symmetry 
is broken. We find that the symmetry breaking  is associated with 
the creation of asymmetry in the charge 
distribution on H atoms upon charging. Removing 
or putting an electron on these molecules without changing the tetrahedral 
geometry may result in charge imbalance on H atoms or 
equal charges on H atoms as in the case of 
CH$_{4}$$^{-}$. This 
charge imbalance or balance  on  H atoms depends on the 
electronic structure of the molecule. In particular, on the nature and 
occupation of the HOMO. 

Using molecular orbital theory\cite{Levin} we discuss the nature of 
HOMO in these  
molecules  and show how it contributes to charge distribution 
among H atoms.  In Fig. 5, we show  the MO construction 
of XH$_{4}$ from 
linear combinations of atomic orbitals. As shown in the figure, there are 
eight MOs, the lower four are bonding states and the upper four are antibonding 
states.  The bonding and antibonding states have  either  A$_{1}$ symmetry 
or  T$_{2}$ symmetry. A$_{1}$ level is non-degenerate while T$_{2}$ level is 
triply degenerate. While the bonding energy level of A$_{1}$ symmetry is 
always lower in energy than T$_{2}$, the order of antibonding levels depends
 on the molecule. For CH$_{4}$ antibonding A$_{1}$ level 
is lower in energy than T$_{2}$ while for SiH$_{4}$, GeH$_{4}$, 
SnH$_{4}$ and PbH$_{4}$ antibonding T$_{2}$ level 
is lower  than A$_{1}$. From the 
figure we can see that the MO with A$_{1}$ symmetry 
is a  linear combination of valance atomic s orbitals of X and H 
atoms.  The MO with T$_{2}$ symmetry is a linear combination of 
valance p orbitals of X and s orbitals of H atoms.  Because of mixing 
between p and s 
character, T$_{2}$ orbitals have strong directionality along H atoms. If 
T$_{2}$ orbital is fully occupied, it has equal weight along 
H atoms and results  
in a uniform charge distribution on H atoms 
as in the case of neutral XH$_{4}$ molecules. If T$_{2}$ orbital is 
incompletely filled with unequal electrons,  
it has unequal weight along H atoms and results in a 
non-uniform charge distribution on H atoms.  

Removing an 
electron from tetrahedral XH$_{4}$ results    
with 5 electrons and the T$_{2}$ orbital is 
incompletely filled. Hence  the charges on H atoms 
of tetrahedral XH$_{4}$$^{+}$ molecules are not equal. This can be seen from  
Fig. 6(a) and (b) which show the charge densities   
of SiH$_{4}$ and SiH$_{4}$$^{+}$ in tetrahedral structure.  We  
see  from the figure that while the charge densities on H atoms of SiH$_{4}$ 
are same, the charge densities on H atoms of SiH$_{4}$$^{+}$ are different. 
 In the case of 
negatively charged 
molecules the HOMO can be of T$_{2}$ or A$_{1}$ 
symmetry. The HOMO of CH$_{4}$$^{-}$ has A$_{1}$ symmetry. 
Since A$_{1}$ is  
a linear combination of valance  s atomic orbitals with equal weights
from hydrogen atoms, 
 the charge distribution is equal on H atoms of  
CH$_{4}$$^{-}$ as shown in Fig. 6(c).  The higher electron 
density around the carbon atom as seen in the figure  can be attributed to 
the higher electronegativity of 
carbon compared to hydrogen.  The HOMO of other 
negatively charged molecules has T$_{2}$ symmetry and is
singly occupied. This 
leads to asymmetric charge distribution on H atoms. Note that since 
A$_{1}$ level in CH$_{4}$$^{-}$ is non-degenerate, CH$_{4}$$^{-}$ does not show the 
Jahn-Teller effect, and also no charge asymmetry. 
On the other hand, if the HOMO has  T$_{2}$ symmetry and is 
not completely  occupied as in the case of SiH$_{4}$$^{+}$,
 the molecule will show the Jahn-Teller effect and the charge asymmetry.  
This implies that the creation of the charge asymmetry and the Jahn-Teller 
effect are intimately connected.

\section{ Conclusions}

We have investigated  the ground state structures of neutral and 
charged XH$_{4}$ molecules using first principle electronic 
structure methods. We find that charging XH$_{4}$ molecules can 
break the symmetry and  change 
their structures drastically. Negatively charged molecules are distorted 
by pushing two hydrogen atoms away from each other while the positively 
charged molecules get distorted with an unusual H-H bond formation except 
for X=C. The calculations done by various methods and the charge density 
analysis confirm the existence of H-H bond. Furthermore,  vibrational 
analysis shows that 
the strength of H-H bond in the positively charged molecule 
is stronger than H$_{2}$$^{+}$ and weaker than H$_{2}$.  We find that 
one vibrational mode of C$_{s}$ symmetric 
XH$_{4}$$^{+}$ molecules has low frequency and corresponds to 
internal rotation of the H-H pair.  Our vibrational analysis indicates 
that the molecules are not transition states.  The fragmentation 
behavior of XH$_{4}$$^{+}$ into XH$_{2}$ and H$_{2}$ shows that the 
stability decreases from CH$_{4}$$^{+}$ to PbH$_{4}$$^{+}$.  

From the 
total energy calculations using LDA, LSD, GGA and GGA-SP 
functionals  we have shown that 
the  ground state electronic structure of a 
charged tetrahedral molecule 
has asymmetric charge density  corresponding to 
integer occupation. We have provided an insight into the 
symmetry breaking and shown that the distortion and 
unusual H-H 
bond formation  can be understood 
easily using simple arguments based on charge asymmetry and electrostatic 
interaction between the atoms. Unequal charges on hydrogen atoms  
give rise to different electrostatic forces on different atoms and lead to 
distortion. On the other hand if the charges are equal on  hydrogen 
atoms, the structural symmetry of the molecule will be preserved. We have 
shown how the charge asymmetry arises due to  character of the HOMO and 
eventually leads to the symmetry breaking.

\begin{acknowledgments}

 It is a pleasure to thank Drs. R. Ramaswamy, D. D. Sarma, 
S. C. Agarwal and N. Sathyamurthy for helpful discussions
and comments. This work was supported by the Department of Science
and Technology, New Delhi via project No. SP/S2/M-51/96.

\end{acknowledgments}

\newpage

\begin{center}
\bf{FIGURE CAPTIONS}
\end{center}
{\bf Fig. 1}. Ground state structures of neutral and charged 
XH$_{4}$ molecules. 
\\
 {\bf Fig. 2} Electron density contour of SiH$_{4}$$^{+}$  in the 
plane of Si and the two bonded H atoms. 
\\
{\bf Fig. 3} The total energy  of SiH$_{4}$$^{+}$, H$_{2}$ 
and H$_{2}$$^{+}$ molecules as 
a function of H-H distance. The total energy of the molecule is given  
with respect its ground state energy. 
\\
{\bf Fig. 4}. (a) The Si-H bond lengths  and  (b) H-H distances
in SiH$_{4}$$^{+}$  as a function of CG steps. The  
H atoms have the same label as those of SiH$_{4}$$^{+}$ in  Fig. 1. 
(c) The sum of energy levels of SiH$_{4}$$^{+}$ as a function of 
CG steps. 
\\
{\bf Fig. 5}. A schematic energy-level diagram showing the construction  
of molecular orbitals  of XH$_{4}$ molecule from atomic orbitals. The 
levels on the extreme left side are the valance 
states of X atoms and the level on the extreme right is the valance state 
of H atom. The levels which are in the center are the molecular states of 
XH$_{4}$. 
\\
{\bf Fig. 6}. The isosurface contour of valance electrons    
of (a) SiH$_{4}$ (b) SiH$_{4}$$^{+}$  and (c) CH$_{4}$$^{-}$ obtained by 
using the Gaussian package.  The value 
of the contour is 
chosen such that one able to distinguish 
the  difference between the charge density on the H atoms.  The densities 
shown in  figure have the contour value of 0.20 for  SiH$_{4}$ and 
SiH$_{4}$$^{+}$  molecules and  0.3  for CH$_{4}$$^{-}$. 

\pagebreak 

Table. I.  Relative energies ( in eV)  of  XH$_{4}$$^{+}$ molecules 
for various symmetries with respect to the ground state structures. The 
results are    
obtained using the Gaussian98 with the BPW91 exchange-correlation 
functional. 

\begin{longtable}[c]{|p{2cm}||p{1cm}|p{1cm}|p{1cm}|p{1cm}|}\hline
	
	&{\sf C$_{2v}$}
	&{\sf D$_{2d}$}
	&{\sf C$_{3v}$}
	&{\sf C$_{s}$}
\\\hline
\hline
	 {\sf CH$_{4}$$^{+}$}
	&{\sf 0.11}
	&{\sf 0}
	&{\sf -}
	&{\sf -}
\\\hline
	 {\sf SiH$_{4}$$^{+}$}
	&{\sf -}
	&{\sf 0.73}
	&{\sf 0.73}
	&{\sf 0}
\\\hline
	 {\sf GeH$_{4}$$^{+}$}
	&{\sf 0.83}
	&{\sf 0.94}
	&{\sf 0.87}
	&{\sf 0}
\\\hline
	 {\sf SnH$_{4}$$^{+}$}
	&{\sf 1.36}
	&{\sf 1.46}
	&{\sf 1.35}
	&{\sf 0}
\\\hline
	 {\sf PbH$_{4}$$^{+}$}
	&{\sf 1.80}
	&{\sf 1.91}
	&{\sf 1.72}
	&{\sf 0}
\\\hline
\end{longtable}

\vspace{1cm} 

Table. II Properties of H-H bond calculated using the Gaussian98 package with 
the BPW91 exchange-correlation functional.  

\begin{longtable}[c]{|p{5cm}||p{1.5cm}|p{1.5cm}|p{1.5cm}|p{1.5cm}|p{1.5cm}|
p{1.5cm}|}\hline
	
	 {\sf  }
	&{\sf SiH$_{4}$$^{+}$}
	&{\sf GeH$_{4}$$^{+}$}
	&{\sf SnH$_{4}$$^{+}$}
	&{\sf PbH$_{4}$$^{+}$}
	&{\sf H$_{2}$}
	&{\sf H$_{2}$$^{+}$}
\\\hline
\hline
	 {\sf H-H bond length( $\AA$ )}
	&{\sf 0.80}
	&{\sf 0.78}
	&{\sf 0.77}
	&{\sf 0.76}
	&{\sf 0.75}
	&{\sf 1.12}
\\\hline
	 {\sf H-H frequency ( cm$^{-1}$ )}
	&{\sf 3598}
	&{\sf 3911}
	&{\sf 4061}
	&{\sf 4134}
	&{\sf 4349}
	&{\sf 1976}
\\\hline
	 {\sf H-H force constant( eV/$\AA$)}
	&{\sf 7.69}
	&{\sf 9.08}
	&{\sf 9.79}
	&{\sf 10.15}
	&{\sf 11.23}
	&{\sf 2.32}
\\\hline
\end{longtable}

\vspace{1cm}

Table. III. The energy required to fragment XH$_{4}$$^{+}$  into 
XH$_{2}$$^{+}$ and H$_{2}$. The results are obtained using 
the Gaussian98 package with the BPW91 exchange-correlation functional.  

\begin{longtable}[c]{|p{3cm}||p{3cm}|p{3cm}||p{3cm}|}\hline
	 {\sf Reactant} 
	&{\sf product 1 }
	&{\sf product 2 }
	&{\sf Fragmentation}
\\
        &
        &
        &{energy( eV )} 
\\\hline 
\hline 
        
	 {\sf CH$_{4}$$^{+}$}
	&{\sf CH$_{2}$$^{+}$}
	&{\sf H$_{2}$}
	&{\sf 3.03}
\\\hline
	 {\sf SiH$_{4}$$^{+}$}
	&{\sf SiH$_{2}$$^{+}$}
	&{\sf H$_{2}$}
	&{\sf 0.69}
\\\hline
	 {\sf GeH$_{4}$$^{+}$}
	&{\sf GeH$_{2}$$^{+}$}
	&{\sf H$_{2}$}
	&{\sf 0.42}
\\\hline
	 {\sf SnH$_{4}$$^{+}$}
	&{\sf SnH$_{2}$$^{+}$}
	&{\sf H$_{2}$}
	&{\sf 0.28}
\\\hline
	 {\sf PbH$_{4}$$^{+}$}
	&{\sf PbH$_{2}$$^{+}$}
	&{\sf H$_{2}$}
	&{\sf 0.21}
\\\hline
\end{longtable}
\pagebreak 

Table. IV The calculated vibrational frequencies $\nu$( in cm$^{-1}$ ) 
and force constant k( in eV/$\AA$ ) of neutral and
charged XH$_{4}$ molecules using the Gaussian98 package with the BPW91
exchange-correlation potential.

\begin{longtable}[c]{|p{1.0cm}||p{1cm}|p{1cm}||p{1cm}|p{1cm}||p{1cm}|
p{1cm}||p{1cm}|p{1cm}||p{1cm}|p{1cm}|p{1cm}|}\hline

\hline
\hline
&
\multicolumn{9}{c}{ Neutral molecules}
&
\\\hline
&
\multicolumn{1}{c}{CH$_{4}$}
&
&
\multicolumn{1}{c}{SiH$_{4}$}
&
&
\multicolumn{1}{c}{GeH$_{4}$}
&
&
\multicolumn{1}{c}{SnH$_{4}$}
&
&
\multicolumn{1}{c}{PbH$_{4}$}
&
\\\hline
\hline
	 {\ }
	&{\  $\nu$} 
	&{\  k }
	&{\   $\nu$}
	&{\  k }
	&{\  $\nu$}
	&{\  k }
	&{\  $\nu$}
	&{\  k }
	&{\  $\nu$}
	&{\  k }
\\\hline
	 {\  1}
	&{\  1298}
	&{\  1.17}
	&{\  879}
	&{\  0.50}
	&{\  788}
	&{\  0.38}
	&{\  676}
	&{\  0.28}
	&{\  644}
	&{\  0.25}
\\\hline
	 {\  2}
	&{\  1298}
	&{\  1.17}
	&{\  880}
	&{\  0.50}
	&{\  788}
	&{\  0.38}
	&{\  676}
	&{\  0.28}
	&{\  644}
	&{\  0.25}
\\\hline
	 {\  3}
	&{\  1298}
	&{\  1.17}
	&{\  881}
	&{\  0.50}
	&{\  788}
	&{\  0.38}
	&{\  676}
	&{\  0.28}
	&{\  644}
	&{\  0.25}
\\\hline
	 {\  4}
	&{\  1521}
	&{\  1.37}
	&{\  949}
	&{\  0.53}
	&{\  901}
	&{\  0.48}
	&{\  735}
	&{\  0.32}
	&{\  703}
	&{\  0.29}
\\\hline
	 {\  5}
	&{\  1521}
	&{\  1.37}
	&{\  950}
	&{\  0.54}
	&{\  901}
	&{\  0.48}
	&{\  735}
	&{\  0.32}
	&{\  703}
	&{\  0.29}
\\\hline
	 {\  6}
	&{\  2975}
	&{\  5.26}
	&{\  2175}
	&{\  2.81}
	&{\  2058}
	&{\  2.51}
	&{\  1856}
	&{\  2.05}
	&{\  1769}
	&{\  1.86}
\\\hline
	 {\  7}
	&{\  3090}
	&{\  6.19}
	&{\  2191}
	&{\  2.97}
	&{\  2071}
	&{\  2.59}
	&{\  1856}
	&{\  2.07}
	&{\  1786}
	&{\  1.91}
\\\hline
	 {\  8}
	&{\  3090}
	&{\  6.19}
	&{\  2191}
	&{\  2.97}
	&{\  2072}
	&{\  2.59}
	&{\  1857}
	&{\  2.07}
	&{\  1786}
	&{\  1.91}
\\\hline
	 {\  9}
	&{\  3090}
	&{\  6.19}
	&{\  2191}
	&{\  2.97}
	&{\  2072}
	&{\  2.59}
	&{\  1857}
	&{\  2.07}
	&{\  1787}
	&{\  1.91}
\\\hline
\hline
&
\multicolumn{9}{c}{ Positively charged molecules }
&
\\\hline
 {\  }  
&
\multicolumn{1}{c}{CH$_{4}$$^{+}$}  
&
&
\multicolumn{1}{c}{SiH$_{4}$$^{+}$}  
&
&
\multicolumn{1}{c}{GeH$_{4}$$^{+}$}  
&
&
\multicolumn{1}{c}{SnH$_{4}$$^{+}$}  
&
&
\multicolumn{1}{c}{PbH$_{4}$$^{+}$}  
&
\\\hline
\hline
	 {\  }
	&{\  $\nu$}
	&{\  k }
	&{\  $\nu$}
	&{\  k }
	&{\  $\nu$}
	&{\  k }
	&{\  $\nu$}
	&{\  k }
	&{\  $\nu$}
	&{\  k }
\\\hline
	 {\  1}
	&{\  470}
	&{\  0.15}
	&{\  256}
	&{\  0.04}
	&{\  136}
	&{\  0.01}
	&{\  46}
	&{\  0.00}
	&{\  -104}
	&{\  0.01}
\\\hline
	 {\  2}
	&{\  471}
	&{\  0.15}
	&{\  616}
	&{\  0.25}
	&{\  510}
	&{\  0.16}
	&{\  380}
	&{\  0.09}
	&{\  316}
	&{\  0.06}
\\\hline
	 {\  3}
	&{\  1022}
	&{\  0.78}
	&{\  680}
	&{\  0.28}
	&{\  532}
	&{\  0.17}
	&{\  394}
	&{\  0.09}
	&{\  339}
	&{\  0.07}
\\\hline
	 {\  4}
	&{\  1253}
	&{\  0.93}
	&{\  766}
	&{\  0.35}
	&{\  542}
	&{\  0.18}
	&{\  430}
	&{\  0.11}
	&{\  396}
	&{\  0.09}
\\\hline
	 {\  5}
	&{\  1420}
	&{\  1.20}
	&{\  850}
	&{\  0.45}
	&{\  658}
	&{\  0.26}
	&{\  495}
	&{\  0.15}
	&{\  425}
	&{\  0.11}
\\\hline
	 {\  6}
	&{\  2642}
	&{\  4.26}
	&{\  1077}
	&{\  0.69}
	&{\  781}
	&{\  0.37}
	&{\  666}
	&{\  0.27}
	&{\  631}
	&{\  0.24}
\\\hline
	 {\  7}
	&{\  2758}
	&{\  4.52}
	&{\  2164}
	&{\  2.83}
	&{\  1957}
	&{\  2.29}
	&{\  1729}
	&{\  1.78}
	&{\  1506}
	&{\  1.35}
\\\hline
	 {\  8}
	&{\  2851}
	&{\  5.32}
	&{\  2242}
	&{\  3.15}
	&{\  2046}
	&{\  2.54}
	&{\  1807}
	&{\  1.96}
	&{\  1623}
	&{\  1.57}
\\\hline
	 {\  9}
	&{\  2852}
	&{\  5.32}
	&{\  3598}
	&{\  7.69}
	&{\  3911}
	&{\  9.08}
	&{\  4061}
	&{\  9.79}
	&{\  4134}
	&{\  10.15}
\\\hline
\pagebreak
\caption[ ]{ Continued}\\\hline
\hline
&
\multicolumn{9}{c}{ Negatively charged molecules }
&
\\\hline
 {\  }  
&
\multicolumn{1}{c}{CH$_{4}$$^{-}$}  
&
&
\multicolumn{1}{c}{SiH$_{4}$$^{-}$}  
&
&
\multicolumn{1}{c}{GeH$_{4}$$^{-}$}  
&
&
\multicolumn{1}{c}{SnH$_{4}$$^{-}$}  
&
&
\multicolumn{1}{c}{PbH$_{4}$$^{-}$}  
&
\\\hline
\hline
	 {\  }
	&{\  $\nu$}
	&{\  k }
	&{\  $\nu$}
	&{\  k }
	&{\  $\nu$}
	&{\  k }
	&{\  $\nu$}
	&{\  k }
	&{\  $\nu$}
	&{\  k }
\\\hline
	 {\  1}
	&{\  1133}
	&{\  0.88}
	&{\  690}
	&{\  0.30}
	&{\  614}
	&{\  0.23}
	&{\  491}
	&{\  0.15}
	&{\  347}
	&{\  0.07}
\\\hline
	 {\  2}
	&{\  1133}
	&{\  0.88}
	&{\  759}
	&{\  0.36}
	&{\  729}
	&{\  0.33}
	&{\  619}
	&{\  0.23}
	&{\  469}
	&{\  0.13}
\\\hline
	 {\  3}
	&{\  1133}
	&{\  0.88}
	&{\  790}
	&{\  0.38}
	&{\  768}
	&{\  0.35}
	&{\  645}
	&{\  0.25}
	&{\  673}
	&{\  0.27}
\\\hline
	 {\  4}
	&{\  1372}
	&{\  1.12}
	&{\  955}
	&{\  0.58}
	&{\  884}
	&{\  0.48}
	&{\  784}
	&{\  0.37}
	&{\  675}
	&{\  0.27}
\\\hline
	 {\  5}
	&{\  1372}
	&{\  1.12}
	&{\  1000}
	&{\  0.59}
	&{\  934}
	&{\  0.52}
	&{\  794}
	&{\  0.37}
	&{\  692}
	&{\  0.29}
\\\hline
	 {\  6}
	&{\  2537}
	&{\  3.82}
	&{\  1323}
	&{\  1.09}
	&{\  1099}
	&{\  0.73}
	&{\  1067}
	&{\  0.68}
	&{\  766}
	&{\  0.35}
\\\hline
	 {\  7}
	&{\  2631}
	&{\  4.55}
	&{\  1429}
	&{\  1.22}
	&{\  1323}
	&{\  1.04}
	&{\  1245}
	&{\  0.92}
	&{\  1124}
	&{\  0.75}
\\\hline
	 {\  8}
	&{\  2631}
	&{\  4.55}
	&{\  1958}
	&{\  2.32}
	&{\  1786}
	&{\  1.91}
	&{\  1607}
	&{\  1.54}
	&{\  1472}
	&{\  1.29}
\\\hline
	 {\  9}
	&{\  2631}
	&{\  4.55}
	&{\  1962}
	&{\  2.37}
	&{\  1804}
	&{\  1.96}
	&{\  1615}
	&{\  1.56}
	&{\  1489}
	&{\  1.32}
\\\hline
\end{longtable}

\pagebreak

Table. V Relative energies( in eV ) of SiH$_{4}$$^{+}$ 
and CH$_{4}$$^{+}$ in tetrahedral structure 
for various electronic configuration with respect 
to charge asymmetric configuration, using different exchange-correlation 
functionals.  

\begin{longtable}[c]{|p{2cm}|p{4.5cm}||p{1.5cm}|p{1.5cm}|}\hline
         {\ }
	&{\ configuration }
	&{\ SiH$_{4}$$^{+}$}
	&{\ CH$_{4}$$^{+}$}
\\\hline
         {\ LDA }
	&{\centering {\ 2, 2, 1}}
	&{\ 0.000}
	&{\ 0.000}
\\

         {\ } 
	&{\centering {\ $\frac{5}{3}$, $\frac{5}{3}$, $\frac{5}{3}$}}
	&{\ 0.001}
	&{\ 0.001}
\\\hline
         {\ GGA} 
	&{\centering {\ 2, 2, 1}}
	&{\ 0.000}
	&{\ 0.000}
\\

         {\ } 
	&{\centering {\ $\frac{5}{3}$, $\frac{5}{3}$, $\frac{5}{3}$}}
	&{\ 0.030}
	&{\ 0.000}
\\\hline
         {\ LSD} 
	&{\centering {\ 1$\uparrow$, 1$\uparrow$, 1$\uparrow$( 1$\downarrow$, 
1$\downarrow$, 0$\downarrow$)}}
	&{\ 0.000}
	&{\ 0.000}
\\
         {\ } 
	&{\centering {\ $\frac{5}{3}$$\uparrow$, $\frac{5}{3}$$\uparrow$, 
$\frac{5}{3}$$\uparrow$( $\frac{5}{3}$$\downarrow$, $\frac{5}{3}$$\downarrow$, 
$\frac{5}{3}$$\downarrow$)}}
	&{\ 0.139}
	&{\ 0.296}
\\
         {\ }  
	&{\centering {\ 1$\uparrow$, 1$\uparrow$, 1$\uparrow$( 
$\frac{2}{3}$$\downarrow$, $\frac{2}{3}$$\downarrow$, 
$\frac{2}{3}$$\downarrow$)}}
	&{\ -0.028}
	&{\ 0.036}
\\\hline
         {\ GGA-SP } 
	&{\centering {\ 1$\uparrow$, 1$\uparrow$, 1$\uparrow$( 
1$\downarrow$, 1$\downarrow$, 0$\downarrow$)}}
	&{\ 0.000}
	&{\ 0.000}
\\
         {\ } 
	&{\centering {\ $\frac{5}{3}$$\uparrow$, $\frac{5}{3}$$\uparrow$,
 $\frac{5}{3}$$\uparrow$ ( $\frac{5}{3}$$\downarrow$, 
$\frac{5}{3}$$\downarrow$,$\frac{5}{3}$$\downarrow$)}}
	&{\ 0.215}
	&{\ 0.333}
\\
         {\ } 
	&{\centering {\ 1$\uparrow$, 1$\uparrow$, 1$\uparrow$ ( 
$\frac{2}{3}$$\downarrow$, $\frac{2}{3}$$\downarrow$,
$\frac{2}{3}$$\downarrow$)}}
	&{\ 0.070}
	&{\ 0.105}
\\\hline
\end{longtable}


\begin{thebibliography}{99}
\bibitem{Steep} We have also looked at the evolution of SiH$_{4}$$^{+}$ 
structure using the steepest descent method. Although the convergence is 
slow, the results are similar to those shown in Fig. 4.
\bibitem{Pop}  The numbers shown in the text are 
obtained from the natural  population analysis using BPW91 exchange correlation 
potential. The numbers obtained by the Mulliken analysis are somewhat different
but show the same trend and thus do not change our conclusions.
\end{thebibliography}
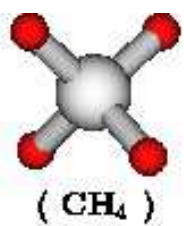
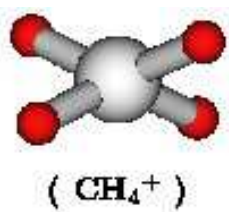
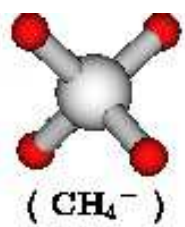
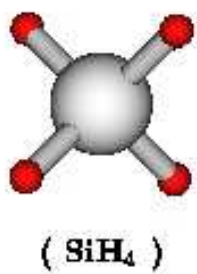
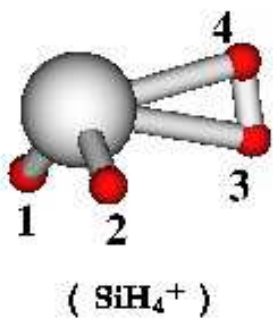
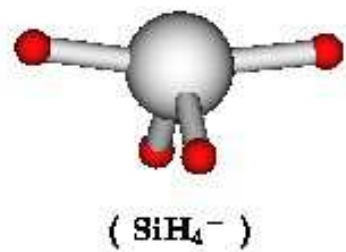
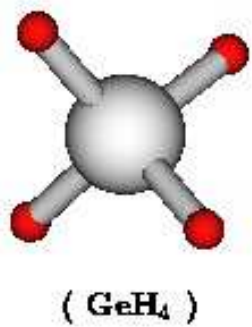
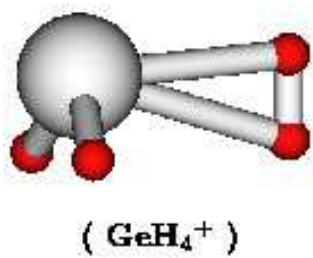
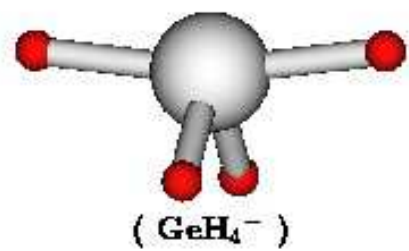
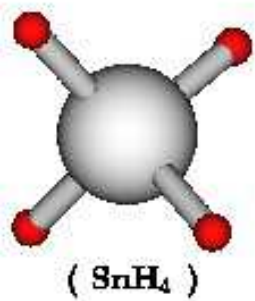
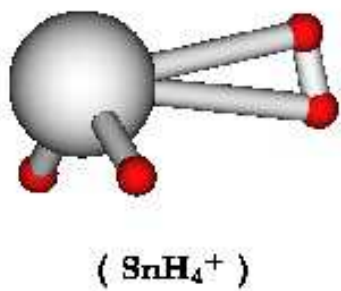
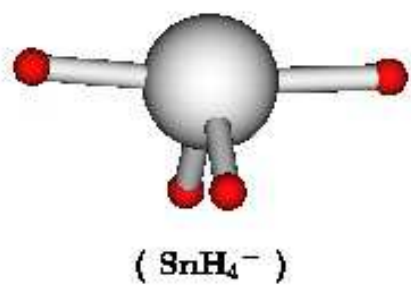
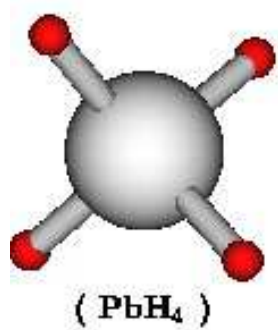
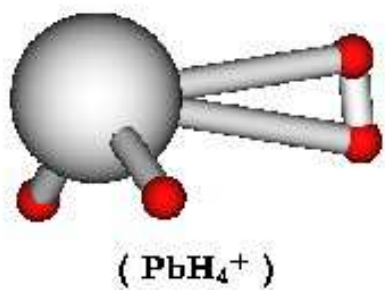
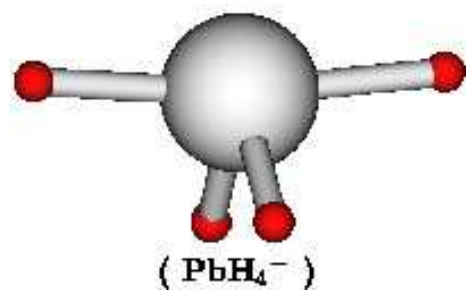
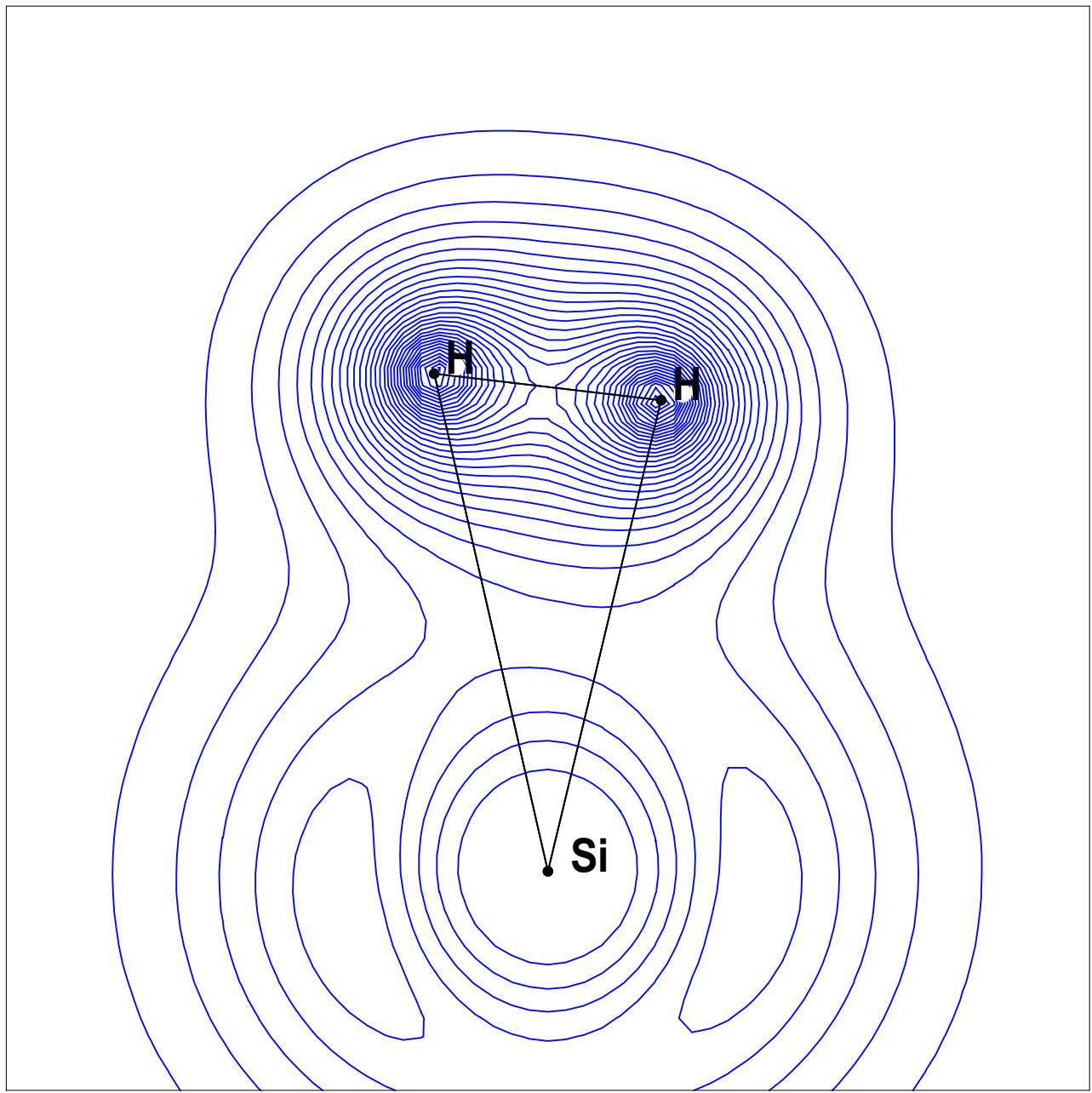
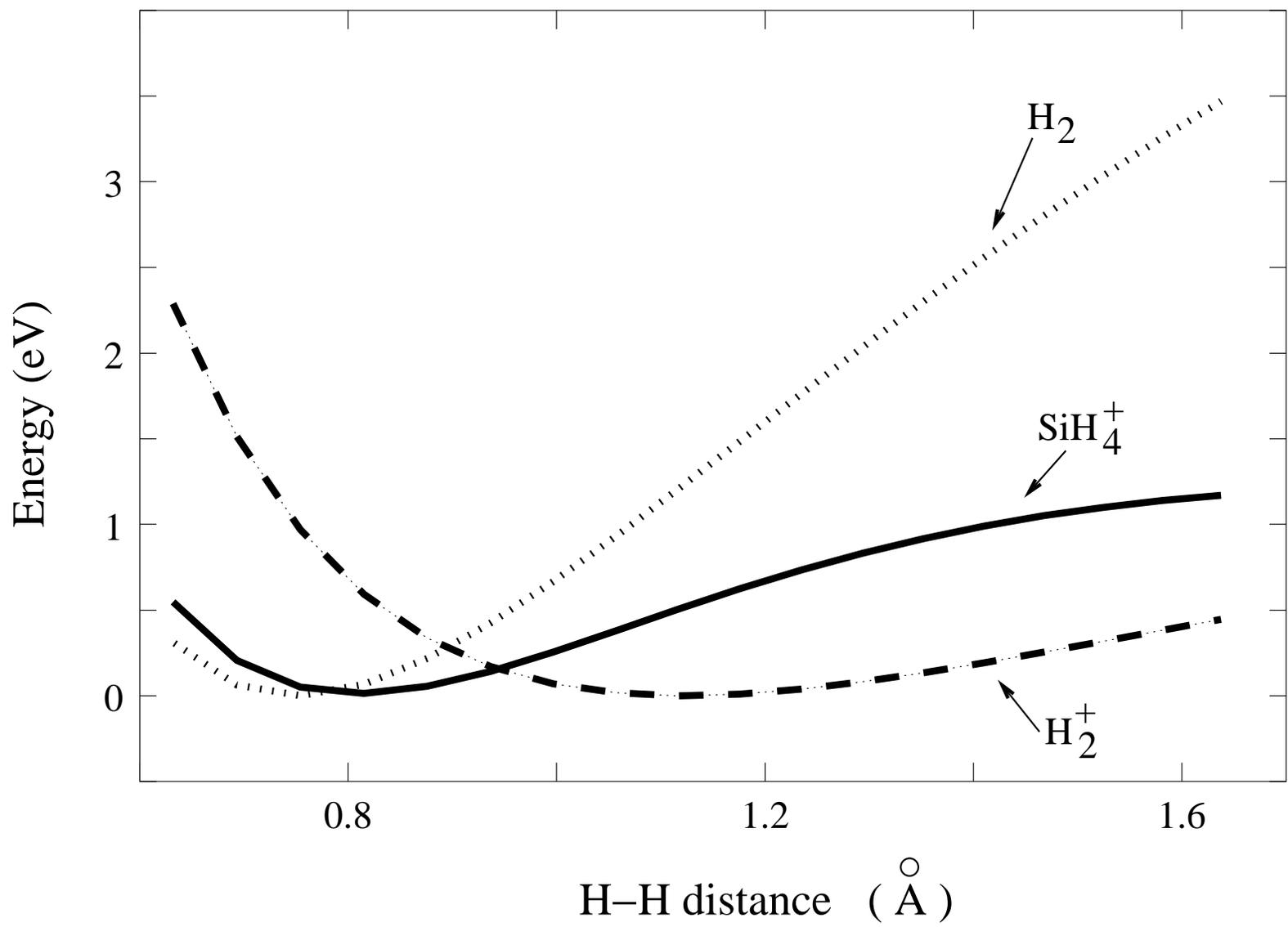
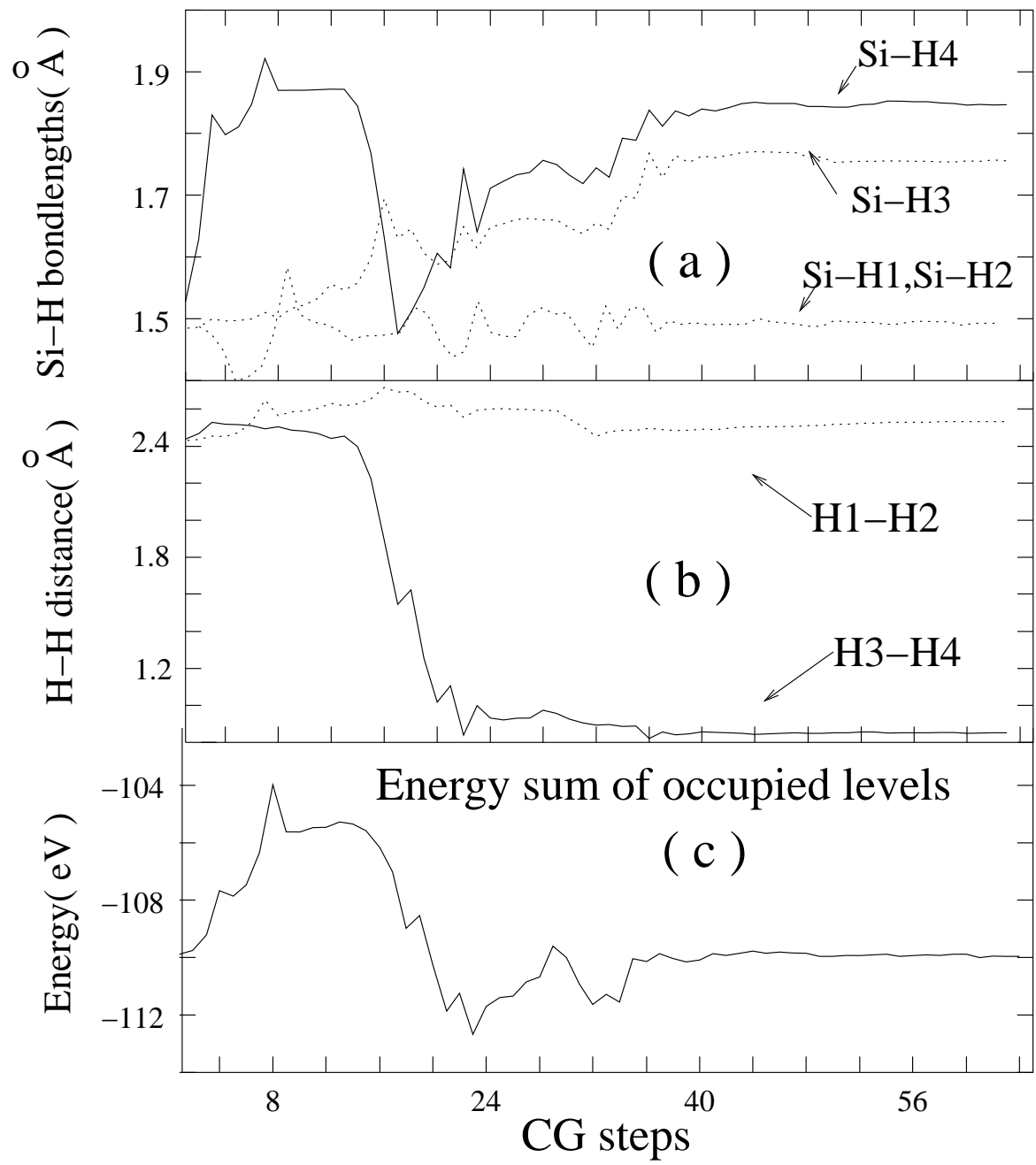
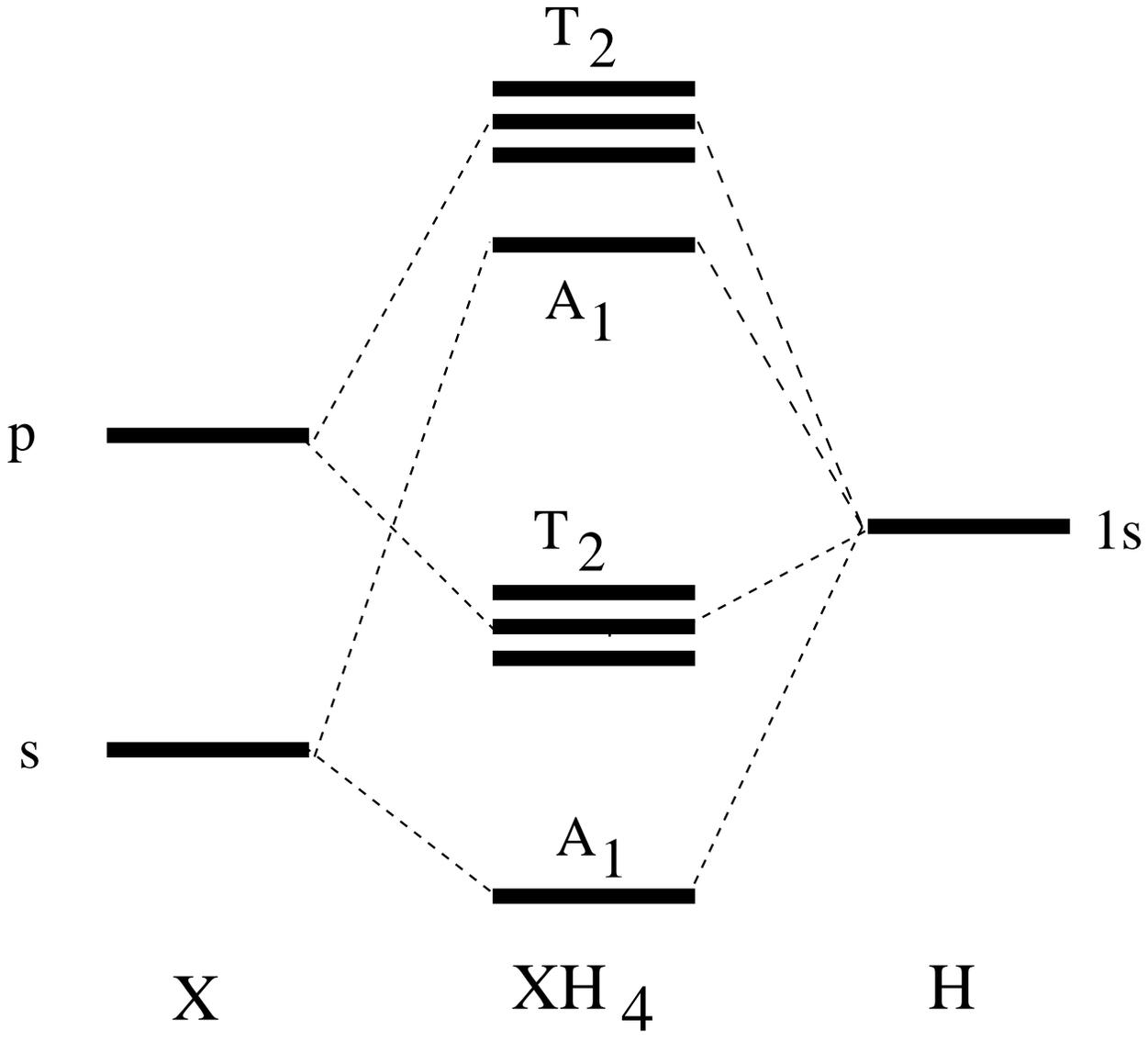
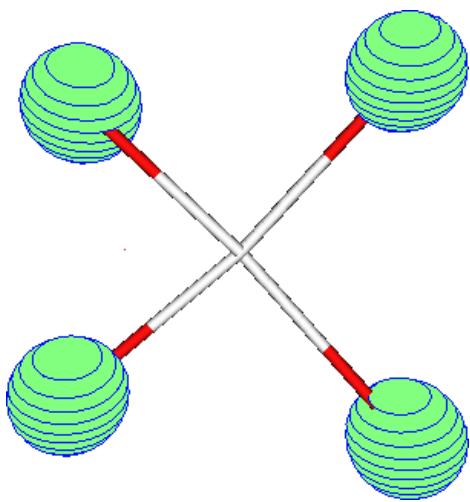
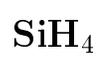
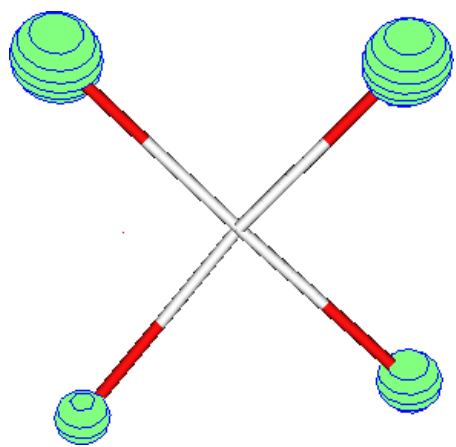
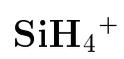
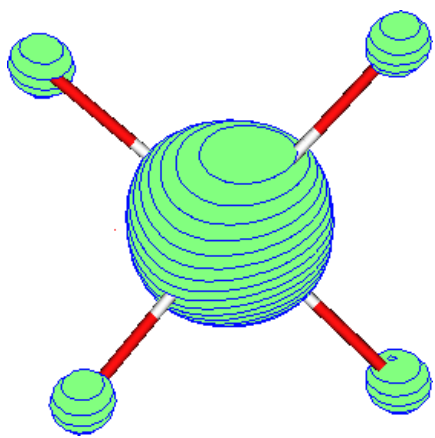
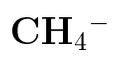
\end{document}